# Rapid Activation of Non-Oriented Mechanophores via Shock Loading and Spallation

Brenden W. Hamilton, Alejandro Strachan[1]

School of Materials Engineering and Birck Nanotechnology Center, Purdue University, West Lafayette, Indiana, 47907 USA

## Abstract

Mechanophores, stimuli-responsive molecules that respond chromatically to mechanochemical reactions, are important for understanding the coupling between mechanics and chemistry as well as in engineering applications. However, the atomic-level understanding of their activation originates from gas phase studies or under simple linear elongation forces directly on molecules or polymer chains containing mechanophores. The effect of many-body distortions, pervasive in condensed-phase applications, is not understood. Therefore, we performed large-scale molecular dynamics simulations of a PMMA-spiropyran co-polymer under dynamic mechanical loading and studied the activation of the mechanophore under various conditions from dynamical compression to tension during unloading. Detailed analysis of the all-atom MD trajectories shows that the mechanophore blocks experience significant many-body intra-molecular distortion that can significantly decrease the activation barrier as compared to when deformation rates are slow relative to molecular relaxation timescales. We find that the reactivity of mechanophores under material compression states is governed by many-body effects of intra-molecular torsions, whereas under tension the reactions are governed by tensile stresses.

---

[1] Corresponding author: strachan@purdue.edu

# Introduction

Mechanochemistry, the use of mechanical loads to trigger or influence chemistry, can alter chemical kinetics and decomposition paths in covalent systems [1]. These processes have been widely studied for single molecules and gas-phase systems both experimentally and computationally. [2] Such processes and a general understanding of mechanochemistry are currently applied in industrial applications of various chemicals and materials [3,4]. In general, condensed phase mechanochemistry extends to a broad range of chemical events such as phase transitions in carbon [5], planetary collisions [6–8], and shockwave initiation of high explosives [1,9–12].

Mechanophores are stimuli-responsive materials that undergo chemical reactions, such as isomerization or bond scission, under mechanical stimuli, and this results in a macroscopic, chromatic response, such as color change or fluorescence [4]. The broad range of applications of these materials results in significant interest from both the engineering and basic sciences communities. Hence, significant experimental and computational work has been devoted to increasing our understanding of the governing physics [13–16]. Such a knowledgebase may, in turn, enable the design of mechanophores for a variety of applications such as flexible electronics [17], aerospace, [18] energy infrastructure, [19] catalysis [20], material strengthening [21], and the release of other molecules under controlled conditions [22].

Computational efforts have significantly increased our understanding of mechanochemistry within condensed matter systems. Motivated by experiments using atomic force microscopy to measure covalent bond strength [23], quantum chemistry (QC) and molecular dynamics (MD) methods have proven invaluable in understanding mechanochemical activation and reactions [2,24]. QC and MD methods have been the dominant workhorses for these studies. They helped elucidate the activation processes in mechanophores [13,14,25] and those driving the reduction in strength in knotted and entangled polymer systems [26–29], assess extension and bond breakage in metal-organic junctions under strain [30–32], and explore the stress states associated with protein folding [33]. Often these types of studies, especially those for mechanophores, use special purpose methods such as steered MD [34–38] and the constrained geometries simulate external force (CoGEF) method [39].

These methods are often applied to mechanophores as single truncated systems in the gas phase and neglect a variety of many-body effects incurred in the condensed matter state, especially under shock loading or other high strain rate events. Recent computational studies showed that dynamic strain-induced chemistry in molecular materials can occur at significantly faster rates than under thermodynamically identical, but equilibrium, conditions [9]. Additionally, it has been shown that shockwave-induced shear bands will react orders of magnitude faster than non-sheared sections of the material [10]. The mechanochemical effects have been recently attributed to large intra-molecular deformations induced by the high strain rates and significant plastic flow associated with dynamical loading [40–42]. These deformations are significant distortions of a many-body nature; primarily out-of-plane bends and torsional rotations induced in the molecule. Such intramolecular strains can persist well into the timescales of chemical initiation and are known to influence the reaction kinetics and reaction paths [43–47]. The recently developed many-body steered MD method (MBsMD) has been used to show that these complex deformations generally both increase reactivity and alter the 1$^{st}$ step reaction pathways [48]. Specifically, torsional shears applied to the central C-O bond of a spiropyran molecule showed up to a 45% reduction in activation barrier for strain-induced isomerization as compared to simple pulling forces.

Spiropyran, which is used in this work, is a mechano-chromic molecule that undergoes a stress-induced 6π electrocyclic ring-opening reaction via bond scission of the C-O bond, resulting in merocyanine [13]. The activation of these mechanophore reactions is greatly influenced by the properties of the polymer system in which it is incorporated [49]. Conversely, independent of the polymer system, be it glassy or elastomeric, the external forces resolved along the reaction coordinate of the mechanophore must drive the strain energy above a threshold value to induce reaction [50,51]. The activation of these materials can be strain rate-dependent [52], and at slow strain rates, such as in a tensile test, the deformation is distributed rather uniformly throughout the material. Under these conditions, systems typically need to approach macroscopic material failure to induce significant activation of the mechanophore. This is more easily achieved in chains/mechanophores aligned with the macroscopic pulling direction where the macroscopic applied force couples better to the action of elongating the mechanophore. Additionally, compression experiments in PMMA with spiropyran mechanophores showed that the highest levels of activation were localized where Poisson effects maximize the tensile stresses in the lateral directions [13,53] and material experiences a high deviatoric state.

Activation can be greatly influenced by the various environmental factors that affect chain mobility [54]. In general, the response of the mechanophore in condensed matter systems can be greatly influenced by the complex dynamics of polymer chains under dynamic loading, which become increasingly more complex under shock loading [55]. Here, we use dynamical compression and expansion, with timescales much shorter than those associated with chain re-alignment, to study spiropyran activation. Our results show the significance of many-body molecular distortions on activation and the need to consider more than simple elongation forces when studying mechanophore reactions, as well as condensed matter chemistry in general.

For mechanophores embedded in an amorphous polymer system, not all local stresses will be uniaxial and aligned with the active bond. Resolved shears, local collisions, entanglements, and crosslinks can result in some measure of a many-body strain such as out-of-plane bends or torsional rotations. Here we utilize reactive MD simulations of shock loading on a spiropyran-PMMA system in which the sample geometry is designed to go through a progression of compressive and tensile loads that can then induce spall and failure. We show that these loads induce a variety of many-body deformations in the spiropyran block of the chains and that both compression and tension lead to rapid activation. This is compared to a compression-only baseline result. The many-body deformation (strain) and complex stress-velocity states of the activated mechanophores is assessed.

## Methods

Simulations are conducted using all-atom MD via the LAMMPS software [56]. The ReaxFF reactive forcefield was used with the ReaxFF-LG parametrization [57] that has been previously used to explore polymer crosslinking [58], graphitization of carbon [59], and reactions in energetic materials [60,61]. A mechanophore system (MP) consisting of a co-polymer of PMMA and spiropyran was built using PolymerModeler [62], an open, online tool available on nanoHUB. The systems are built as PMMA/indolinobenzospiropyran copolymers with the spiropyran as the central monomer block with 40 PMMA monomers on either end (81 monomers total per chain). Using a continuous configurational bias Monte Carlo method [63], we built an initial condensed matter system with 625 chains for a total of 920,000 atoms.

The MP system was initially relaxed using the nonreactive Dreiding forcefield [64] at 800 K in a fully periodic system keeping a 4:1:1 ratio between cell parameters under isobaric, isothermal (NPT) conditions for 100 ps. The system was then cooled to 300 K with a cooling rate of 100 K/ns, at ambient pressure, forcing the cell aspect ratio to remain the same. The glass transition obtained from these simulations, 450 K, was comparable to pure PMMA under the same conditions [65].

The 300 K structure was further relaxed using the reactive force field ReaxFF and divided into a projectile and target sections, see Figure 1, to enable dynamical loading simulations via high-velocity impacts. Two gaps were introduced in the system, by deleting whole chains whose centers of mass fall within a predetermined region, along the long cell direction. The first gap is at the cell boundary to break periodicity, and the second at 1/3 distance into the system. This results in two pairs of free surfaces and a left-hand slab 15nm in length and a right-hand slab 32.5 nm in length separated by a 10 nm gap, as shown in Figure 1. The right-hand side system contains 364 chains, and the left 148 chains for a final total of 512 total chains.

Ballistic impact simulations [66] were performed with impact velocities ranging from 2.0 to 4.0 km/s. Two distinct systems were used. In the "two-slab system" as shown in Figure 1, slabs were initialized with impact velocities of equal and opposite vectors, where the collision creates two shockwaves of equal strength propagating in opposite directions into the "flyer" and "target". As these compressive waves reach a free surface, they become tensile in nature and a rarefaction fan propagates back into the sample. This setup was designed to generate a wide range of compressive and tensile states, which sample a wide range of deviatoric stress states, subjecting the mechanophore to various dynamical loads and, hypothetically, induce many-body intra-molecular strains. A second set of simulations were carried out to explore the response of the system under sustained shock compression. In this setup, the smaller, left slab (projectile) is held perfectly rigid and is impacted in the reverse-ballistic manner by the large slab leading to a single compressive shock. When the shock reaches the free surface at the far end of the target slab, shock absorbing boundary conditions (SABCs) [40,67,68] are applied to keep the system in the compressed state indefinitely. We will refer to this second setup as "sustained shock" and the first one as "shock and spall".

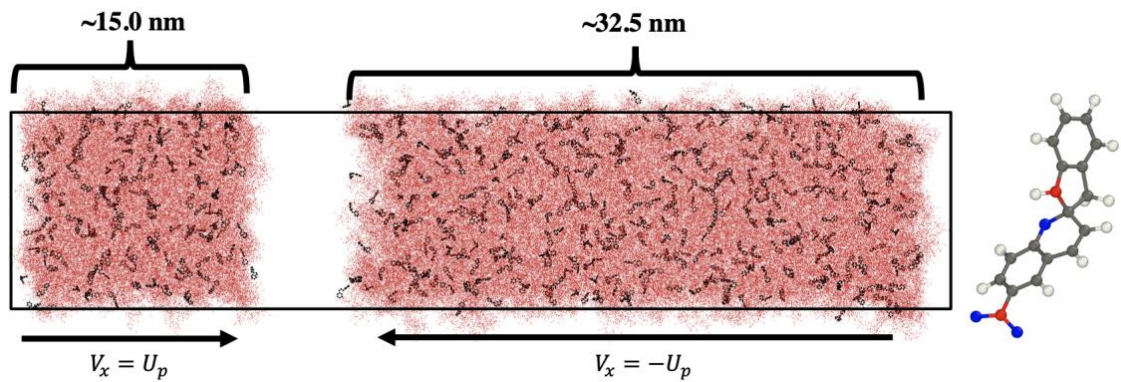

*Figure 1: Set-up for shock simulations of the Spiropyran + PMMA system. Atoms colored in red are PMMA backbone carbons, atoms in black are Spiropyran ring atoms. For the 'one-slab' system, the left piece is kept perfectly rigid to approximate an infinitely massive piston.*

## Results and Discussion

Figure 2 shows x-t diagrams along the shock direction colored by local particle velocity along the shock direction, temperature, and pressure tensor component in the shock direction ($P_{xx}$, the negative of the stress component $\sigma_{xx}$) for the shock & spall system with a $u_p$ of 4 km/s. The plots show the propagation of the two compressive shockwaves and rarefaction fans after the interaction of the shocks with the free surfaces. The leftward moving wave reaches the free surface just after 2 ps, creating a rarefaction fan that propagates in the positive x direction at a faster velocity than the shock (shock velocity is the inverse of the slope in x-t diagrams). The rightward moving shock reaches the other free surface at ~4.5 ps, forming a second rarefaction fan and expanding the material outwards. The two rarefaction waves meet at about 5ps, creating a state of high tensile stress that leads to spall failure. Figure 2 (b) shows that the highest temperature occurs at the impact plane and that the adiabatic cooling upon release/expansion at the free surfaces is on the order of ~200 K. Figures 2 (c) and (d) show $P_{xx}$ at two different scales: the full range of values (panel c from -5 to 35 GPa) and a small range around zero to better see tensile states (panel d from -2 to 2 GPa). Figure 2 (c) shows a shock pressure of about 35 GPa with an almost instantaneous rise. The rarefaction fan forms at the surface, resulting in a nearly instantaneous drop in pressure at the surface, the fan then continuously expands as it propagates across the material. In the last time frame, the fan has expanded such that the pressure drop takes 1.5 ps. Figure 2 d) shows that the rarefaction waves reach tensile stresses of nearly -2 GPa on fast rise times which results in the spall failure of the material and significant extension.

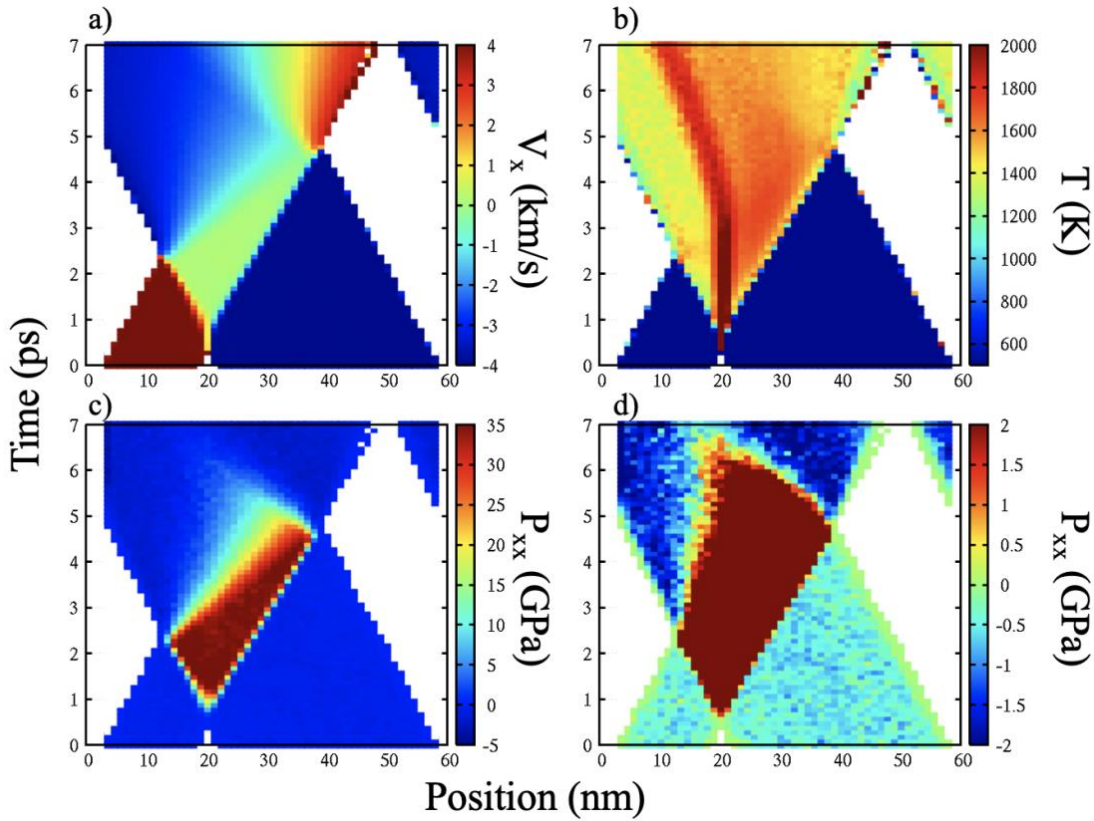

*Figure 2: Shock x-t diagrams for $U_p$ = 4.0 km/s. Panel a is colored by particle velocity (positive is left to right motion), panel b is roto-vibrational temperature, c is pressure and d is pressure with a narrowed color range to better view tensile states.*

Previous work has shown the shock compression of molecular solids can lead to significant intra-molecular strain [40,41,69]. The high pressures of the shock state prevent the molecules from conformationally relaxing in a significant way, and this can accelerate chemical reactions [10,48]. The characterize the level of strain of the mechanophores, we track two measures of intra-molecular strains: the distance between the MP anchor atoms that connect the spiropyran with the PMMA on both ends and the torsional dihedral angle around the C-O bond that breaks during the isomerization reaction. The first measure directly tracks the molecular elongation that is often assumed responsible for inducing the reaction. The second is an intra-molecular rotation that has been traditionally ignored in mechanochemistry studies but has recently been shown to lower the activation barrier of isomerization [48]. Figure 3 shows the distribution of these measures of intramolecular strain under various conditions. The distributions for all MPs in the system 0.2 ps after being shocked (different absolute times for each molecule) are shown as green lines and red lines correspond to the MPs that reacted during the shock and spall simulation (independent of compression or tension), we report the values 0.2 ps prior to reaction. To establish a baseline, these deformations have also been measured in an equilibrated sample under no mechanical load (300 K, 1 atm), see blue lines in Fig. 3, as well as for a periodic system subjected to uniaxial tensile deformation at a strain rate of $1 \times 10^{-9}/s$, averaged over 1.5% and 2.5% strain (brown lines).

The uniaxial quasistatic strain has little effect on the distributions of each deformation, which is not surprising for a glassy PMMA system that will not undergo significant chain re-alignment on such small strains and short timescales. However, under shock compression, which has been shown to significantly increase chain activity directly behind the shockwave [55], the distributions of molecular deformations are significantly broadened. For the distance between anchor atoms, shock compression naturally leads to significant shortening as compared to the equilibrium distribution, as the spiropyran molecules can fold/rotate around the spiro atom. However, this is not an entirely fair comparison as the maximum compression of the shockwave is ~50% strain. Interestingly, there is also a slight expansion of the distribution towards longer distances, which may be a result of spiropyran monomers being flattened and therefore extended in the transverse direction to the shockwave, or effects from thermal expansion, or rapid molecular re-alignments [55]. As hypothesized, the compression of the MPs also leads to significant many-body distortions, originating from the rotation around the spiro atom, extending the tail of the dihedral angle distribution by ~50°. Lastly, the deformations of the MPs that react appear to follow the distributions of the shocked states, however, the tails are relatively oversampled. This shows that the MPs that react on shock rise and rarefaction timescales are not limited to those with significant intra-molecular strain, but the many-body strain does influence the reactivity, increasing its likelihood, in agreement with previous results in many-body mechanochemistry [48].

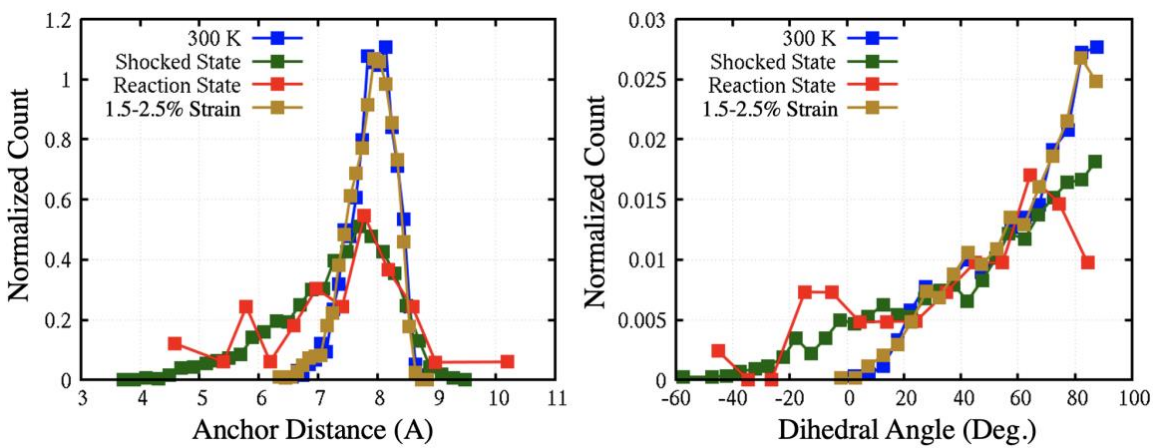

*Figure 3: Distributions of spiropyran molecular deformations. Anchor distance is the distance between the two spiropyran carbon atoms bonded to the PMMA backbone carbons. Dihedral angle is the angle formed from a N-C-O-C atoms where the central carbon is the spiro atom and the central bond is the C-O bond that breaks during reaction.*

Figure 4 shows a scatter plot of the anchor distances and dihedral angles for all MPs 0.2 ps after loading (black) and those that react (red) in the shock and spall case for 4 km/s, dashed lines indicate the equilibrium values at 300 K. As described above, shock loading results in large molecular strains that are not just compression/elongation on the structure but also many-body in nature. Interestingly, the state of the MP prior to reaction does not follow the same distribution as the shocked states. MPs experiencing elongation are more likely to react, this is expected. We note that reactions under tension does not require large many-body strains. Quite remarkably, a significant fraction of MPs undergoes reaction under compression, almost invariably with a significant degree of many-body strain. The Pearson correlation coefficients between anchor distance and dihedral angle for the shock and reaction states are -0.069 and 0.373, respectively. The trend of the reacted points show the lack of need for tension with increasing torsion, which was previously shown to lower the activation barrier [48]. While many shocked MPs exist in the compression state (low anchor distance), very few react without significant torsion, and the wide spread of torsion states at small distances for the shocked states shows that this is not just a geometric trend that the molecules rotate when compressed.

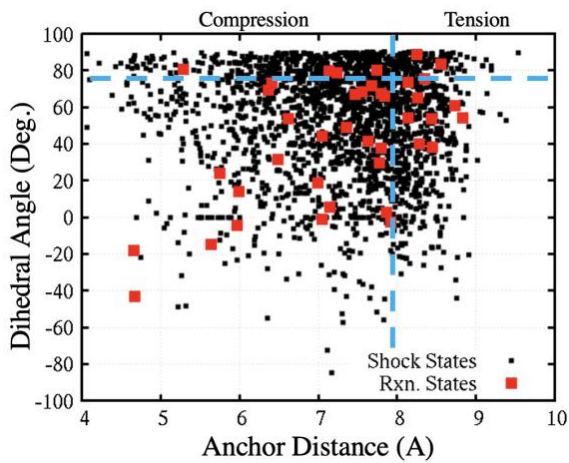

*Figure 4: 2D distribution of the deformations shown in Figure 3. Black dots represent shock states and red dots the reacted states. Blue dashed lines are the equilibrium values for each axis.*

Figure 5 displays the fraction of MPs that have undergone reaction as a function of simulation time for the shock & spall system (left) and the sustained shock system that only experiences compression (right). This is cumulative reactions, however, if a spiropyran molecule finds a path to undergo the reverse reaction, it is no longer counted. While the conditions in the shock and spall case are experienced over much shorter timescales, relative to the sustained shock, the shock and spall cases show considerably more reactivity, which can be attributed to the presence of tensile stress and the transient states during rarefaction.

Focusing on the shock & spall system, the rate of reaction during the early compressive regime show little to no effect from shock strength except for the weakest shock cases (2.0 and 2.5 km/s). This agrees well with the sustained shock results. However, once the first rarefaction wave forms (2-3 ps), there is a strong dependence of activation rate on the shock strength. This tracks with the idea that the rarefaction-induced tension and rate of expansion increase with increasing shock strength. The compressive states most likely lead to more many-body intra-molecular strains like the shear induced rotation around the spiro atom, as the higher density restricts spatial relaxations, in which the level of intra-molecular strain can play just as much of a role in controlling the reactivity. Since the majority of MPs under shock compression are not experiencing local tensile strains, their reactions are likely dominated by many-body effects. Supplemental Materials section SM-1 shows that the torsional deformations shown in Figure 3 are not overly influenced by shock strength in the regimes studied here, which agrees well with the trends in reactivity for pure compression. In these cases, the unreacted MPs may be nearing the physical limit of deformations without breaking bonds, meaning that stronger shocks cannot physically push the molecules further. The reactions at later times are driven by tensile stresses which will more directly correspond to shock strength due to the strong relationship of shock strength and the tensile state achieved via rarefaction.

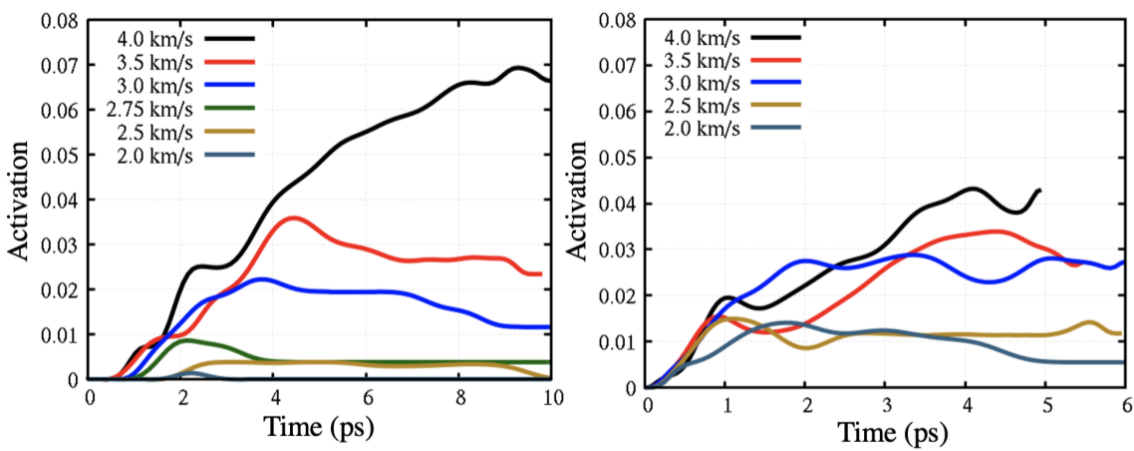

*Figure 5: Fraction of spiropyran molecules with broken C-O bonds (at the spiro atom) across time for all impact velocities.*

To further investigate the local conditions needed to activate MPs, Figure 6 maps all reactions based on their local pressure and its time-derivative right before reaction, for the shock and spall case with impact velocity of 4.0 km/s. Points are colored by the center of mass velocity of the corresponding MPs in the shock direction. Pressure was determined by averaging the local atomic stresses in a sphere with a radius of 1.5 nm, centered at the spiro atom. Time derivatives of the stress were calculated from 0.3 to 0.1 ps prior to reaction.

Based on the pressure and its time derivative, the reaction states of the MPs can be grouped into four main categories. States with positive pressure time derivative and nonzero pressures are reactions that occur during loading, with the velocity distinguishing the leftward or rightward moving shock. Reactions at high pressure and near-zero slope occur under compression, following the passage of the shock, and should typically be near zero velocity as well. A few reactions are observed during the unloading process, they correspond to nonzero pressure and negative slopes. While these states experience an expanding system, they have yet to reach a tensile state. The last cluster of reactions corresponds to near-zero slope and negative pressures, labeled as unloaded and spall. The majority of these states have large, negative particle velocities, meaning that the rarefaction fan has projected the molecules forward in an expansion. Supplemental Materials section SM-2 zooms in on these states and shows that they have pressures between -1 and -2 GPa.

In summary, the MPs react under a variety of conditions, including compression where chemistry is driven by local shears strains and many-body deformations, during spall/expansion, which is driven by the tensile states, and in the transients in between the two. Roughly half of the reactions occur during transient states, either loading or unloading, in which glassy polymers are known to rapidly evolve under shock loading [55], and the remaining half of the reactions happen under either compression and tension when pressure/strain is more constant. However, due to the relevant length and timescales of both compression and tension, as well as the disparate energy states of the two, it is difficult to resolve a general efficiency of one type of loading against the other.

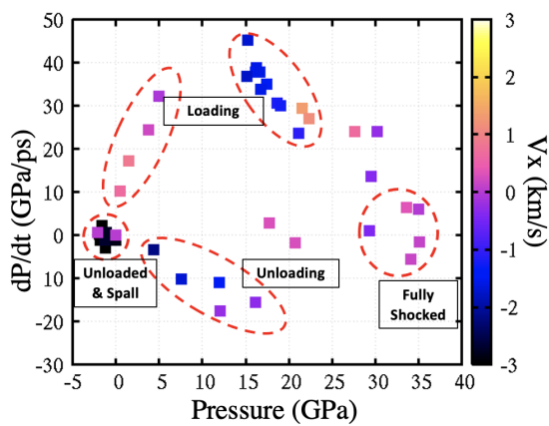

*Figure 6: Parametric plot of the pressure and pressure time derivative directly before C-O bond scission for reaction spiropyran in the $U_P$ = 4.0 km/s case. Color is velocity in the shock direction, which is zero in the compressed state.*

## Conclusions

In this work, shock simulations were conducted using an all-atom reactive representation of a PMMA-spiropyran copolymer. The system geometry was designed to induce significant compressive and tensile stresses leading to spall, which would also incur significant resolved shear stresses. These complex and highly deviatoric stress states lead to significant intra-molecular strain in the mechanophore, which is characterized via the distance between the anchor atoms of the mechanophore (an effective molecular length) and the dihedral angle around the central torsion of the atom, which contains the C-O bond that breaks during isomerization. During rarefaction and spall of the material during a shock with up=4 km/s, tensile stresses of nearly 2 GPa are reached, while the compressive shock nears 40 GPa of pressure. Both induce significant mechanophore activation, with the tensile/spall activation being much more shock strength-dependent than under pure compression. Reactivity in the compression and tension systems is shown to be governed by shock strength, while the reactivity in the compression only states is not shock strength effects, showing the compression parts of reactivity are most likely governed by many-body effects. Lastly, an analysis of the stress-velocity states during activation reveals that activation not only occurs during the compressed and strained states, but during loading and unloading, as well as in the relaxed system, post spall, where the material has begun to fail and does not hold a local stress. These results show the significant need for a better understanding of how deviatoric stress states and intra-molecular strain states can alter condensed matter chemistry, especially that of mechanochemistry.

## Acknowledgments

This work was primarily supported by the US Office of Naval Research, Multidisciplinary University Research Initiatives (MURI) Program, Contract: N00014-16-1-2557. Program managers: Chad Stoltz and Kenny Lipkowitz. We acknowledge computational resources from nanoHUB and Purdue University through the Network for Computational Nanotechnology.

## Supplemental Materials

### SM-1

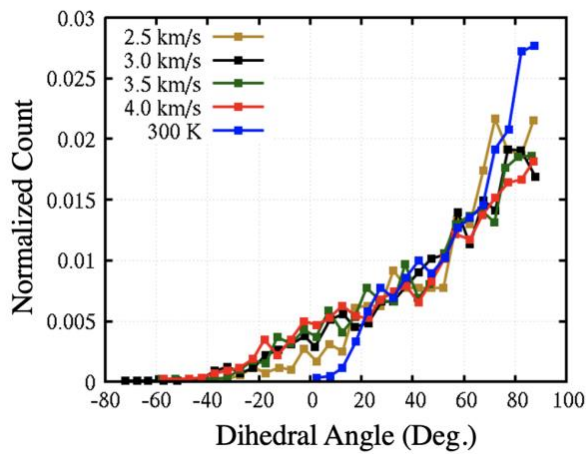

SM Figure 1: Dihedral angle distributions for shock mechanophores and the relaxed system as 300 K.

## SM-2

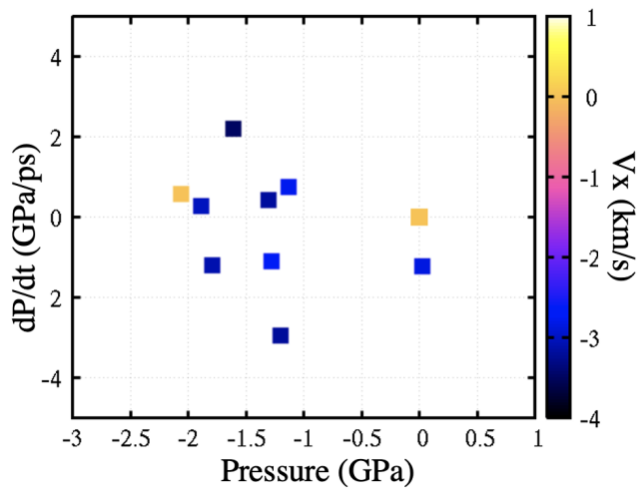

SM Figure 2: Pressure, Pressure slope, and particle velocity data for mechanophores that react under tension.